\documentclass[lettersize,journal]{IEEEtran}
\usepackage[utf8]{inputenc}
\usepackage{bm, cite, float, amsmath, amssymb, amsthm}
\usepackage{amsmath,amsfonts}
\usepackage{algorithmicx}
\usepackage{algorithm}
\usepackage{algpseudocode}
\usepackage{array}
\usepackage[caption=false,font=normalsize,labelfont=sf,textfont=sf]{subfig}
\usepackage{textcomp}
\usepackage{stfloats}
\usepackage{url}
\usepackage{xcolor}
\usepackage{verbatim}
\usepackage{graphicx}
\usepackage{cite,citesort}
\begin{document}

\title{Security–Spectral Efficiency Tradeoff in STAR-RIS RSMA: A Max–Min Fairness Framework}

\author{Huiyun~Xia,~
        Yijie~Mao, ~\IEEEmembership{Member,~IEEE,}
        Sai~Xu, ~\IEEEmembership{Member,~IEEE,}          
        Shuai~Han, ~\IEEEmembership{Senior Member,~IEEE,}        
        and Hongbo Zhu,~\IEEEmembership{Member,~IEEE}
\thanks{This work was supported in part by the Natural Science Research Start-up Foundation of Recruiting Talents of Nanjing University of Posts and Telecommunications under Grant XK0020924009 and in part by the Natural Science Foundation of Jiangsu Province Higher Education Institutions under Grant 25KJB510019 (Corresponding author: Sai Xu).}
\thanks{ Huiyun~Xia and Hongbo Zhu are with the Jiangsu Key Laboratory of Wireless Communications, Nanjing University of Posts and Telecommunications, Nanjing 210003, China (emails: xiahy2024@njupt.edu.cn; zhb@njupt.edu.cn).}
\thanks{Yijie Mao is with the School of Information Science and Technology, ShanghaiTech University, Shanghai, China (email: maoyj@shanghaitech.edu.cn).}
\thanks{S. Xu is with the Department of Electronic and Electrical Engineering, University College London, WC1E 7JE, London, UK.  (e-mail: sai.xu@ucl.ac.uk). }
\thanks{Shuai~Han is with the School of Electronics and Information Engineering, Harbin Institute of Technology, Harbin, 150001, P. R. China (email: hanshuai@hit.edu.cn). }}

\markboth{Journal of \LaTeX\ Class Files,~Vol.~14, No.~8, August~2021}%
{Shell \MakeLowercase{\textit{et al.}}: A Sample Article Using IEEEtran.cls for IEEE Journals}


\maketitle

\begin{abstract}
Simultaneously transmitting and reflecting reconfigurable intelligent surfaces (STAR-RISs) enable full-space coverage but also expose wireless transmissions to security from multiple spatial directions. This paper investigates a STAR-RIS-assisted secure RSMA system where both internal and external eavesdroppers may coexist in the transmission and reflection regions. In such a scenario, the RSMA common stream simultaneously serves legitimate users, impairs external eavesdroppers, and avoids assisting internal eavesdroppers, leading to a challenging trade-off between spectral efficiency and confidentiality.
To address this issue, we formulate a max–min fairness problem under secrecy constraints and develop an iterative algorithm to jointly optimize transmit beamforming and STAR-RIS phase shifts. 
Simulation results demonstrate that the proposed scheme improves spectral efficiency while maintaining confidentiality.

\end{abstract}

\begin{IEEEkeywords}
Rate-splitting multiple access, STAR-RIS, max-min fairness, physical layer security.
\end{IEEEkeywords}

\section{Introduction}
\IEEEPARstart {S}{imultaneously} transmitting and reflecting reconfigurable intelligent surfaces (STAR-RISs) have emerged as an advanced extension of conventional RIS technology, enabling full-space wireless coverage by jointly controlling transmitted and reflected signals \cite{Xu2021STAR-RIS}. 
This capability provides additional degrees of freedom to significantly improve spectral efficiency. 
However, the omnidirectional operation of STAR-RISs also broadens the spatial region where confidential messages may be intercepted, posing more severe eavesdropping threats compared to half-space RISs \cite{Zhou2024STARjamming}. 
These challenges require effective beamforming strategies that can simultaneously sustain high spectral efficiency and enhance communication confidentiality.
Rate splitting multiple access (RSMA) offers a promising solution by flexibly splitting user messages into common and private parts, and achieving a favorable balance between spectral efficiency and confidentiality \cite{Mao2022survey}. Through tailored beamforming, the common stream can also act as deliberate interference to external eavesdroppers, reinforcing physical layer security by enlarging the channel disparity between legal users and wiretappers \cite{Xia2024sRSMA}.
These features make RSMA particularly attractive for STAR-RIS-assisted networks, where internal and external eavesdroppers may simultaneously appear on both the transmission and reflection sides.

Although STAR-RIS-aided secure RSMA designs have attracted growing attention, existing studies remain limited.
Most prior  works only consider external wiretappers located on a single side of the STAR-RIS \cite{Zhang2024secureSTAR,Hashempour2024secureSTAR,XIAO2023PLS}, whose insights cannot be directly extended to a practical $360^\circ$ coverage scenario where multiple eavesdroppers may exist on both sides.
Besides, the common stream must simultaneously deliver information to legal users, serve as artificial noise to impair external eavesdroppers on both sides, and avoid inadvertently aiding internal eavesdroppers. 
This requires a delicate trade-off between confidentiality and spectral efficiency, which has been insufficiently explored, particularly in the presence of internal eavesdroppers.
The above gaps motivate a comprehensive security-aware RSMA framework that fully leverages the propagation control of STAR-RIS to safeguard communications against both internal and external threats on both sides.
The contributions of this paper are:
1) We investigate the STAR-RIS-assisted secure RSMA framework in a more practical scenario where both internal and external eavesdroppers coexist in the transmission and reflection regions. 
2) To achieve balance between confidentiality and spectral efficiency, an iterative algorithm is developed to jointly optimize beamforming and phase shifts, achieving max-min fairness under stringent secrecy constraints.
3) Numerical results validate that the proposed design significantly enhances spectral efficiency while preserving confidentiality. The performance gain remains stable with increasing STAR-RIS elements but diminishes as eavesdroppers’ channel quality improves.

\section{System Model}
Consider a downlink STAR-RIS-aided RSMA system where an $N_{\rm T}$-antenna transmitter serves $K$ single-antenna users $\mathcal{K} \buildrel\Delta\over= \{1,...,K\}$ via an $N_{\rm S}$-element STAR-RIS. Users are divided into the reflection and  transmission space, denoted by $\mathcal{K}_r$ and $\mathcal{K}_t=\mathcal{K}\backslash\mathcal{K}_r$ ($l\in\{r,t\}$).
The transmitter sends $K$ confidential messages, each intended for UE-$k$ and kept secrecy from both (i) internal eavesdroppers (legal users decoding non-intended messages), and (ii) external eavesdroppers (non-authenticated receivers $J\in\mathcal{J} \buildrel\Delta\over= \{1,...,J\}$) which reside on both sides, denoted by $\mathcal{J}_t=\{1,2,...,J_t\}$, $\mathcal{J}_r=\mathcal{J}\backslash\mathcal{J}_t$.
Perfect CSI is assumed \cite{Xie2025STAR} and direct links are blocked \cite{Hashempour2024secureSTAR}. The STAR-RIS operates in energy splitting mode for full-space coverage \cite{Xu2021STAR-RIS}.

A 1-layer RSMA scheme \cite{Mao2022survey} is employed, where each message is split into a common and private part. 
All common parts are combined and encoded as a common stream $s_\mathrm{c}$ decoded by all users, while each private part is independently encoded into a private stream $s_k$ for UE-$k$ only.  
Let ${\bf{s}} \buildrel\Delta\over= [s_{\mathrm{c}}, s_1, s_2, . . . , s_K]^{\rm{T}}$ with $\mathrm{E}[\bf{s}\bf{s}^{\rm{H}}] = \bf{I}$. 
The transmit signal is
$\mathbf{x}=\mathbf{W}\mathbf{s}=\mathbf{w}_{\mathrm{c}} s_{\mathrm{c}}+\sum\nolimits_{k\in\mathcal{K}}\mathbf{w}_k s_k$, where ${\bf W} \buildrel\Delta\over= [{\bf w}_{\mathrm{c}}, {\bf w}_1, {\bf w}_2, . . . , {\bf w}_K]$ satisfies ${\rm{tr}}({\bf W}{\bf W}^{\rm{H}}) \le P_{\rm max}$.  ${\bf w}_{\mathrm{c}}, {\bf w}_k \in {\mathbb C}^{N_t\times 1}$ are the precoders for $s_{\mathrm{c}}$ and $s_k$. 
$P_{\rm max}$ is the transmit power limit. 
Each STAR-RIS element $n_{\rm s}\in{\mathcal N}_{\rm S}\buildrel\Delta\over=\{1,2,...,N_{\rm S}\}$ has coefficient $l_{n_{\rm s}}=(\beta_{l,{n_{\rm s}}}e^{j\theta_{l,{n_{\rm s}}}})$ with $0<\beta_{l,{n_{\rm s}}}<1$, $\beta_{r,{n_{\rm s}}}^2+\beta_{t,{n_{\rm s}}}^2=1$ and $\theta_{l,{n_{\rm s}}}\in[0,2\pi)$.
Channels from transmitter to STAR-RIS, from STAR-RIS to UE-$k$, and STAR-RIS to Eve-$j$ are denoted by $\mathbf H\in{\mathbb C}^{N_{\rm S}\times N_{\rm T}}$, ${\mathbf h}_k \in{\mathbb C}^{N_{\rm S}\times 1}$, and ${\mathbf g}_j \in{\mathbb C}^{N_{\rm S}\times 1} $,  respectively.

Each user first decodes the common stream by treating private streams as interference, then applies successive interference cancellation (SIC) and decodes its intended private stream. The signal to interference plus noise ratios (SINRs) of $s_{\rm c}$ and $s_k$ at UE-$k$ in the reflection or transmission region are
\begin{equation}\label{eq4_2_2}
\begin{aligned}
{\gamma^{l}_{{\rm{c}},k}} = \frac{{|{\bf{h}}_k^{\rm{H}}{\bf\Theta}^{l}{\mathbf H}{{\bf{w}}_{\rm{c}}}{|^2}}}{I   + \sigma^2},   {\gamma^{l}_{{\rm{p}},k}} = \frac{{|{\bf{h}}_k^{\rm{H}}{\bf\Theta}^{l}{\mathbf H}{{\bf{w}}_k}{|^2}}}{I- {|{\bf{h}}_k^{\rm{H}}{\bf\Theta}^{l}{\mathbf H}{{\bf{w}}_k}{|^2}  + \sigma^2}},
\end{aligned}
\end{equation}
where $I={\sum_{i \in {\cal K}} {|{\bf{h}}_k^{\rm{H}}{{\bf\Theta}^{l}}{\mathbf H}{{\bf{w}}_i}{|^2}}}$,  ${\bf\Theta}^l={\rm diag}(\beta_{l,1}e^{j\theta_{l,1}},...,\beta_{l,N_{\rm S}}e^{j\theta_{l,N_{\rm S}}}) \in{\mathbb C}^{N_{\rm S}\times N_{\rm S}}$ for $l\in\{r,t\}$ is the transmission/reflection coefficient matrix, and 
$\sigma^2$ is the noise variance.
To ensure decodability of $W_{\mathrm{c}}$, the transmission rate $R_{\mathrm{c}}$ should satisfy $R_{\mathrm{c}}\leq\min_{k \in \mathcal{K}} R_{{\mathrm{c}},k}$ with $\sum_{k\in\mathcal{K}}C_k=R_{\mathrm{c}}$, where $C_k$ is the common rate allocated to UE-$k$. 
The total rate of UE-$k$ is $R_{k,{\rm tot}} = C_k+R_{{\mathrm{p}},k}$.

To prevent a legal user from acting as an internal eavesdropper to decode other users' private streams after its own by applying one more layer of SIC, the SINR of any stream $s_i$ at UE-$k$ should remain below the decoding threshold $r_0$:
\begin{equation}\label{gamma_k_i}
\begin{aligned}
r_0\ge\gamma _{k\rightarrow i} &= \frac{{|{\bf{h}}_{k}^{\rm{H}}{{\bf\Theta}^{l}}{\mathbf H}{{\bf{w}}_i}{|^2}}}{{\sum_{k' \in {{\mathcal K}\backslash\{k,i\}}} | {\bf{h}}_{k}^{\rm{H}}{{\bf\Theta}^{l}}{\mathbf H}{{\bf{w}}_{k'}}{|^2}  + \sigma^2}}.
\end{aligned}
\end{equation}
Equation \eqref{gamma_k_i} is imposed to enforce the receiver capability so that  each private stream is exclusively decodable only by its intended user and remains undecodable by any others.

For an external eavesdropper Eve-$j$, the common stream serves as artificial noise. To guarantee security, the SINRs at Eve-$j$ are constrained by the decoding threshold $r_{\rm E}$:
\begin{equation}\label{gamma_c_e}
\begin{aligned}
r_{\rm E} \geq \gamma _{{\rm{c}},j}^{\rm E} &= \frac{{|{\bf{g}}_{j}^{\rm{H}}{{\bf\Theta}^{l}}{\mathbf H}{{\bf{w}}_{\rm{c}}}{|^2}}}{{\sum_{k \in {\cal K}} | {\bf{g}}_{j}^{\rm{H}}{{\bf\Theta}^{l}}{\mathbf H}{{\bf{w}}_k}{|^2}  + \sigma _{\rm E}^2}},
\end{aligned}
\end{equation}

\begin{equation}\label{gamma_p_e}
\begin{aligned}
r_{\rm E} \geq\gamma _{k,j}^{\rm E} &= \frac{{|{\bf{g}}_{j}^{\rm{H}}{{\bf\Theta}^{l}}{\mathbf H}{{\bf{w}}_k}{|^2}}}{{\sum_{k' \in {\tilde {\mathcal K}\backslash\{k\}}} | {\bf{g}}_{j}^{\rm{H}}{{\bf\Theta}^{l}}{\mathbf H}{{\bf{w}}_{k'}}{|^2}  + \sigma_{\rm E}^2}},
\end{aligned}
\end{equation}
where $\tilde {\mathcal K}\buildrel\triangle\over ={\mathcal K}\cup\{{\rm c}\}$.

\section{Problem Formulation and Optimization}
\subsection{Problem Formulation}
We aim to maximize the minimum legitimate user rate while ensuring each confidential message remains exclusively decodable only by its intended receiver. The max-min fairness problem is formulated as
\begin{subequations}\label{P_1}
\begin{align}
\max _{\mathbf{c, P}}  \quad&  \min_{k \in \mathcal{K}} \quad R_{k,{\rm tot}}  \tag{\ref{P_1}}\\
\text { s.t. } 
& {\operatorname{Tr}({\bf{W}}{{\bf{W}}^{\rm{H}}}) \le {P_{max}}}, \label{P_1_c4}\\
& \beta_{r,{n_{\rm s}}}^2+\beta_{t,{n_{\rm s}}}^2=1, \label{power_conserve}\\
& 0<\beta_{t,{n_{\rm s}}},\beta_{r,{n_{\rm s}}}<1, \forall n_{\rm s}\in\mathcal{N}_{\rm S}, \label{beta_constraint}\\
& \theta_{t,{n_{\rm s}}},\theta_{r,{n_{\rm s}}}\in[0,2\pi], \label{theta_constraint}\\
& {{\bf{c}} \ge {\bf{0}}}, \label{P_1_c5} \\
& \nonumber \eqref{gamma_k_i}-\eqref{gamma_p_e},
\end{align}
\end{subequations}
where $\mathbf{c}\buildrel\Delta\over=\{C_1,...,C_K\}$ in constraint \eqref{P_1_c5} specifies non-negative allocation of the common rate. 

\subsection{Optimization Framework}
To handle the non-convex problem \eqref{P_1}, we reconstruct it via semi-definite relaxation (SDR). Define 
$ {\bf{P}}_{m} \buildrel\Delta\over= {\bf{w}}_m{\bf{w}}_m^{\rm H}$, $m\in\tilde {\mathcal K}$ and ${\mathbf q}_l^{\text H}\buildrel\Delta\over=[\beta_{l,1}e^{j\theta_{l,1}},...,\beta_{l,N_{\rm S}}e^{j\theta_{l,N_{\rm S}}}] \in{\mathbb C}^{1\times N_{\rm S}}$. Thus ${\bf \Theta}^l={\rm diag}({\mathbf q}_l^{\text H})$,  ${\mathbf h}_k^{\text H} {\bf \Theta}^l = {\mathbf q}_l^{\text H}{\rm diag}({\mathbf h}_k^{\text H})$.
Denote ${\mathbf Q}^l\buildrel\Delta\over={\mathbf q}_l{\mathbf q}_l^{\text H}$, ${\bf\Gamma}_k \buildrel\Delta\over= {\rm diag}({\mathbf h}_k^{\text H}){\mathbf H}$. 
Therefore, the quadratic forms can be written as $|{\mathbf h}_k^{\text H}{\bf \Theta}^l{\mathbf H}{\mathbf w}_{m}|^2=\operatorname{Tr}({\bf\Gamma}_k{\mathbf P}_{m}{\bf\Gamma}_k^{\rm H}{\mathbf Q}^l)$, for $m\in\{{\rm  c},k,i\}$, $k\in\mathcal{K}$, $i\in {\mathcal K}\backslash\{k\}$. 
Similarly, 
defining ${\bf\Gamma}_j^{\rm E} \buildrel\Delta\over= {\rm diag}({\mathbf g}_j^{\text H}){\mathbf H}$ yields $|{\mathbf g}_j^{\text H}{\bf \Theta}^l{\mathbf H}{\mathbf w}_{m}|^2=\operatorname{Tr}({\bf\Gamma}_k^{\rm E}{\mathbf P}_{m}{{\bf\Gamma}_k^{\rm E}}^{\rm H}{\mathbf Q}^l)$. 
Therefore, problem \eqref{P_1} is equivalently reconstructed as
\begin{subequations}\label{P_2}
\begin{align}
\max_{\mathbf{c, P}, {\bf \Theta}^{l}}    &\min_{k \in \mathcal{K}} \quad R_{k,{\rm tot}} \tag{\ref{P_2}} \\
\text {s.t.} 
& \frac{1}{r_{0}} \operatorname{Tr}({\bf\Gamma}_k{\mathbf P}_{i}{\bf\Gamma}_k^{\rm H}{\mathbf Q}^l)- \sum\nolimits_{k'\in{\mathcal K}\backslash\{k,i\}}\!\!\operatorname{Tr}({\bf\Gamma}_k{\mathbf P}_{k'}{\bf\Gamma}_k^{\rm H}{\mathbf Q}^l) \nonumber\\
& \le  \sigma^{2}, \quad i\in\tilde{\mathcal K}, \label{P_2_in}\\ 
& \frac{1}{r_{\rm E}} \operatorname{Tr}({\bf\Gamma}_j^{\rm E}{\mathbf P}_{k}{{\bf\Gamma}_j^{\rm E}}^{\rm H}{\mathbf Q}^l) - \sum_{k'\in{\mathcal K}\backslash\{k\}}\operatorname{Tr}({\bf\Gamma}_j^{\rm E}{\mathbf P}_{k'}{{\bf\Gamma}_j^{\rm E}}^{\rm H}{\mathbf Q}^l)\nonumber \\
&\leq \operatorname{Tr}({\bf\Gamma}_j^{\rm E}{\mathbf P}_{\rm c}{\Gamma_j^{\rm E}}^{\rm H}{\mathbf Q}^l) + \sigma_{\rm E}^{2}, \quad j\in{\mathcal J}, \label{P_2_c}\\
& \frac{1}{r_{\rm E}}\!\! \operatorname{Tr}({\bf\Gamma}_j^{\rm E}{\mathbf P}_{\rm c}{{\bf\Gamma}_j^{\rm E}}^{\rm H}{\mathbf Q}^l\!) \!\leq \!\!\!\! \sum_{k'\in\mathcal K}\!\!\operatorname{Tr}({\bf\Gamma}_j^{\rm E}{\mathbf P}_{k'}{{\bf\Gamma}_j^{\rm E}}^{\rm H}{\mathbf Q}^l\!) \!\! + \!\! \sigma_{\rm E}^{2}, \label{P_2_d}\\ 
& \sum\nolimits_{k'\in\tilde{\mathcal K}}{\operatorname{Tr}}({\bf{P}}_i) \le P_{\rm max},\label{P_2_e}\\ 
& {\rm rank}({\bf P}_{i}) = 1, \quad i\in\tilde{\mathcal K}, \label{P_2_f}\\
& {\bf P}_{i}\succeq {{0}}, \quad i\in\tilde{\mathcal K}, \label{P_2_g} \\ 
& \nonumber \eqref{power_conserve}-\eqref{P_1_c5}.
\end{align}
\end{subequations}
 
The common rate is presumed equally allocated to all users, i.e., $C_k=\mu_k R_{\rm c}$ with $\mu_k=\frac{1}{K}$ \cite{hao2020robustsecureRS}. 
Since $R_{k,{\rm tot}}$ is non-convex with respect to $\mathbf{c, P}, {\bf \Theta}^{l}$, 
the slack variables $\beta,\tau$ are introduced to reconstruct problem \eqref{P_2}.
Besides, the $\log$ operation is removed as it does not affect optimality, yielding 
\begin{subequations}\label{SSE_5}
\begin{align}
\max_{\beta_k, \tau_k, \mathbf{P}, {\bf \Theta}^l}  &\quad \min_{k\in\mathcal K} \frac{\tau_k\psi_k({\bf P}, {\bf\Theta}^l)}{\phi_k({\bf P}, {\bf\Theta}^l) } \tag{\ref{SSE_5}} \\
\text { s.t. } 
& \tau_k = \beta^{\mu_k}, \label{SSE_4_a}\\
& \beta-1 \le  \min_{k\in\mathcal K} \frac{\operatorname{Tr}({\bf\Gamma}_k{\mathbf P}_{\rm c}{\bf\Gamma}_k^{\rm H}{\mathbf Q}^l)}{\sum_{k'\in\mathcal K}\operatorname{Tr}({\bf\Gamma}_k{\mathbf P}_{k'}{\bf\Gamma}_k^{\rm H}{\mathbf Q}^l) + \sigma^2}, \label{SSE_4_b}\\
&  \eqref{power_conserve}-\eqref{P_1_c5}, \eqref{P_2_in}-\eqref{P_2_g},  \nonumber
\end{align}
\end{subequations}
where
$\psi_k({\bf P}, {\bf\Theta}^l)\buildrel\Delta\over=\sum_{k'\in\mathcal K}\operatorname{Tr}({\bf\Gamma}_k{\mathbf P}_{k'}{\bf\Gamma}_k^{\rm H}{\mathbf Q}^l) + \sigma^2$, $\phi_k({\bf P},{\bf\Theta}^l)\buildrel\Delta\over=\sum_{k'\in{\mathcal K}\backslash\{k\}}\operatorname{Tr}({\bf\Gamma}_k{\mathbf P}_{k'}{\bf\Gamma}_k^{\rm H}{\mathbf Q}^l) + \sigma^2$.

Applying non-linear fractional programming \cite{Zargari2021MMF},  
problem \eqref{SSE_5} can be equivalently reformulated by fixing a parameter $\lambda$ and introducing an auxiliary variable $t$:
\begin{subequations}\label{SSE_6}
\begin{align}
\max_{{\bf c, P}, {\bf\Theta}^l}  &\quad t \tag{\ref{SSE_6}}\\
\text { s.t. } & t \le \tau_k\psi(\mathbf{P}, {\bf \Theta}^l) - \lambda \phi({\bf P}, {\bf \Theta}^l), \label{SSE_6a} \\
&  \eqref{power_conserve}-\eqref{P_1_c5}, \eqref{P_2_in}-\eqref{P_2_g}, \eqref{SSE_4_a}, \eqref{SSE_4_b}, \nonumber
\end{align}
\end{subequations}

Problem \eqref{SSE_6} remains non-convex due to the coupled variables $\bf c$, $\mathbf P$ and ${\mathbf \Theta}^l$. To solve this, an alternate optimization procedure is employed. 
Initially, the transmission and reflection coefficient matrices are optimized with the remaining variables fixed. 
Subsequently, the auxiliary variables and the beamforming covariance matrix $\bf P$ are jointly optimized using the updated transmission and reflection matrices ${{\bf \Theta}^l}^*$.  The detailed procedures will be given in the following two subsections.

\subsubsection{Optimization of phase shifts ${\bf\Theta}^l$ }
Given the optimized variables $\{\tau_k^{[n-1]}, {\bf P}_m^{[n-1]},m\in\tilde{\mathcal K}\}$ from the $[n-1]$-th iteration, the sub-problem for phase shift optimization is formulated as
\begin{subequations}\label{SSE_8}
\begin{align}
\max_{\mathbf{\Theta}^r, {\mathbf \Theta}^t } & \quad t \tag{\ref{SSE_8}}\\
\text { s.t. } & t \leq \tau_k^{[n-1]} \psi( {\mathbf \Theta}^r, {\mathbf \Theta}^t) - \lambda \phi({\mathbf \Theta}^r, {\mathbf \Theta}^t), \forall k, \label{SSE_8_a}\\
& {\operatorname {Rank}}({\mathbf Q}^l) = 1, l\in\{r,t\}, \label{SSE_8_b}\\
& [{\mathbf Q}^r]_{n_{\rm s},n_{\rm s}} + [{\mathbf Q}^t]_{n_{\rm s},n_{\rm s}} = 1, n_{\rm s}\in\mathcal{N}_{\rm S},\label{SSE_8_c}\\
&  \eqref{beta_constraint}, \eqref{theta_constraint}, \eqref{P_2_in}-\eqref{P_2_d},  \eqref{SSE_4_a}, \eqref{SSE_4_b}. \nonumber
\end{align}
\end{subequations} 
Constraints \eqref{SSE_8_b} and \eqref{SSE_8_c} together ensure \eqref{power_conserve} is satisfied.
$\psi_k( {\mathbf \Theta}^r, {\mathbf \Theta}^t)$ and $\phi_k( {\mathbf \Theta}^r, {\mathbf \Theta}^t)$ are convex with respect to ${\bf\Theta}^l,l\in\{r,t\}$ for fixed ${\mathbf P}_m,m\in\tilde{\mathcal K}$. 
Therefore, with any given $\lambda$ in each iteration of optimizing \eqref{SSE_8},
 constraint \eqref{SSE_8_a} is convex. 
The non-convexity arises solely from the rank-one constraint.
To handle this, \eqref{SSE_8_b} is converted into $||{\mathbf Q}^l||_*-||{\mathbf Q}^l||_2=0$, $\forall l\in\{r,t\}$, where ${||\mathbf Q}^l||_*=\sum_i\delta_i({\mathbf Q}^l)$ and $||{\mathbf Q}^l||_2=\max_i\{\delta_i\}$ are the nuclear and spectral norms, respectively \cite{Zargari2021MMF}.
$\delta_i$ refers to the $i$th largest singular value of the matrix ${\mathbf Q}^l$. 
This is due to the fact that for $\forall {\mathbf Q}^l \in{\mathbb H}^{N_{\rm S}\times N_{\rm S}}$, ${\mathbf Q}^l \succeq 0$, $||{\mathbf Q}^l||_*-||{\mathbf Q}^l||_2 \ge 0$ always holds and the equality can be reached if and only if ${\operatorname{Rank}}({\mathbf Q}^l) = 1$.
The rank-one constraint is handled by exploiting the penalty term, yielding
\begin{subequations}\label{SSE_24}
\begin{align}
\max_{{\mathbf \Theta}^r, {\mathbf \Theta}^t} & \quad t-\rho\sum\nolimits_{l\in\{r,t\}}(||{\mathbf Q}^l||_*-||{\mathbf Q}^l||_2) \tag{\ref{SSE_24}}\\
\text { s.t. } 
&  \eqref{beta_constraint}, \eqref{theta_constraint}, \eqref{P_2_in}-\eqref{P_2_d},  \eqref{SSE_4_a}, \eqref{SSE_4_b}, \eqref{SSE_8_a}, \eqref{SSE_8_c}, \nonumber
\end{align}
\end{subequations} 
where $\rho$ is a penalty factor. 
Nevertheless, the formulated problem in \eqref{SSE_24} is still non-convex due to the difference of convex (d.c.) structure in the objective function. In this respect, a lower bound for $A({\mathbf Q}^l)\buildrel\Delta\over=||{\mathbf Q}^l||_2$ is presented by using its first-order Taylor approximation at the current iterate ${{\mathbf Q}^l}^{[n_1\!]}$:
\begin{align}\label{AQ_apprx}
\widetilde A (\!{\mathbf Q}^l\!)\! =\! ||{{\mathbf Q}^l}^{[n_1\!]}||_2 \!\!+\!\! \operatorname{Tr}\!\!\big( {\bf \bar u}({{\mathbf Q}^l}^{[n_1\!]}){\bf \bar u}({{\mathbf Q}^l}^{[n_1\!]})^{\rm H}({\mathbf Q}^l\!-\!{{\mathbf Q}^l}^{[n_1\!]})\big),
\end{align}
where ${\bf \bar u}({{\mathbf Q}^l})$ is the largest eigenvector of ${{\mathbf Q}^l}$. 
Accordingly, at the $[n_1]$-th iteration, problem \eqref{SSE_8} can be reconstructed by
\begin{subequations}\label{SSE_26}
\begin{align}
\max_{{\mathbf \Theta}^r, {\mathbf \Theta}^t} & \quad t-\rho\sum\nolimits_{l\in\{r,t\}}(||{\mathbf Q}^l||_*-{\widetilde A}({\mathbf Q}^l)) \tag{\ref{SSE_26}}\\
\text { s.t. } 
&  \eqref{beta_constraint}, \eqref{theta_constraint}, \eqref{P_2_in}-\eqref{P_2_d},  \eqref{SSE_4_a}, \eqref{SSE_4_b}, \eqref{SSE_8_a}, \eqref{SSE_8_c}, \nonumber
\end{align}
\end{subequations} 
which is convex and can be efficiently solved.

\subsubsection{Optimization of $\bf c$ and $\{{\bf w}_m,m\in\tilde{\mathcal K}\}$ }
Given ${\bf \Theta}^l$, $\psi_k({\bf P})$ and $\phi_k({\bf P})$ are convex with respect to $\bf P$. The second sub-problem are therefore written as
\begin{subequations}\label{SSE_9}
\begin{align}
\max_{{\bf c, P}}  &\quad t \tag{\ref{SSE_9}}\\
& t \le \tau_k\psi(\mathbf{P}) - \lambda \phi({\bf P}), \label{SSE_9a} \\
\text { s.t. } &  \eqref{P_1_c5}, \eqref{P_2_in}-\eqref{P_2_g}, \eqref{SSE_4_a}, \eqref{SSE_4_b}, \nonumber
\end{align}
\end{subequations}

To enable the SDR approach, we first drop the rank constraint \eqref{P_2_f}. Nonetheless, the coupling among the variables $\tau_k$, $\lambda$, and $\beta_k$ makes constraints \eqref{SSE_9a} and \eqref{SSE_4_b} non-convex. Consequently, these constraints are substituted with their first-order Taylor approximations, which provide convex surrogate functions in a neighborhood of the current iterate. Specifically,
\begin{equation}\label{f_apprx}
\begin{aligned}
&f\left(\tau_k, {\mathbf P}, {\bf\Theta}^l\right)  =\tau_k \psi_k\left({\mathbf P}, {\bf\Theta}^l\right)  \\
& \hspace{0.4cm} \approx f\left(\tau_k^{[n_2-1]}, {\mathbf P}^{[n_2-1]}, {{\bf\Theta}^{[n_2-1]}}^l\right) \\
& \hspace{0.4cm} +\sum_{k' \in {\mathcal K}} \operatorname{Tr}\left(\tau_k^{[n_2-1]} ({\bf\Gamma}_k^{\rm H}{{\bf Q}^{[n_2-1]}}^l{\bf\Gamma}_k)^{\rm T}({\mathbf P}_{k'}-{\mathbf P}_{k'}^{[n_2-1]})\right)\\
& \hspace{0.4cm}+ \sum_{l \in \{r,t\}} \!\!\operatorname{Tr}\left(\tau_k^{[n_2-1]} ({\bf\Gamma}_k{{\bf P}_{k'}^{[n_2-1]}}^l{\bf\Gamma}_k^{\rm H})^{\rm T}({{\bf\Theta}^l}-{{\bf\Theta}^{[n_2-1]}}^l)\right) \\
& \hspace{0.4cm}+ \operatorname{Tr}\left((\sum_{k'\in\mathcal K}\operatorname{Tr}({\bf\Gamma}_k{\mathbf P}_{k'}{\bf\Gamma}_k^{\rm H}{\mathbf Q}^l) + \sigma^2)({\tau_k}-{\tau_k}^{[n_2-1]})\right)\\
& \hspace{0.4cm}\buildrel\Delta\over= {\tilde f}\left(\tau_k^{[n_2]}, {\mathbf P}^{[n_2]}, {{\bf\Theta}^{[n_2]}}^l\right),
\end{aligned} 
\end{equation}
where 
$(\tau_k^{[n_2]}, {\mathbf P}^{[n_2]}, {{\bf\Theta}^{[n_2]}}^l)$ is the optimized parameters obtained from the $[n_2]$-th iteration of the inner-layer optimization for solving problem \eqref{SSE_9}. 
Similarly, \eqref{SSE_4_b} is approximated by
\begin{equation} \label{com_strm_cons}
{\tilde g}\left(\beta_k^{[n_2]}, {\mathbf P}^{[n_2]}, {{\bf\Theta}^{[n_2]}}^l\right) \le \operatorname{Tr}({\bf\Gamma}_k{\mathbf P}_{\rm c}{\bf\Gamma}_k^{\rm H}{{\mathbf Q}^l}^{[n]}), \forall k\in\mathcal{K}, 
\end{equation}
and 
\begin{equation}
\begin{aligned}
& {\tilde g}(\beta_k^{[n_2]}, {\mathbf P}^{[n_2]}, {{\bf\Theta}^{[n_2]}}^l) \approx g(\beta_k^{[n_2-1]}, {\mathbf P}^{[n_2-1]}, {{\bf\Theta}^{[n_2-1]}}^l) \\
&+ \sum_{k' \in {\mathcal K}} \operatorname{Tr}\left(({\beta_k^{[n_2-1]}-1}) ({\bf\Gamma}_k^{\rm H}{{\bf Q}^{[n_2-1]}}^l{\bf\Gamma}_k)^{\rm T}({\mathbf P}_{k'}-{\mathbf P}_{k'}^{[n_2-1]})\right)\\
&+ \sum_{l \in \{r,t\}} \operatorname{Tr}(({\beta_k^{[n_2-1]}-1}) ({\bf\Gamma}_k{{\bf P}_{k'}^{[n_2-1]}}^l{\bf\Gamma}_k^{\rm H})^{\rm T} ({{\bf\Theta}^l}-{{\bf\Theta}^{[n_2-1]}}^l)) \\
&+ \operatorname{Tr}((\sum\nolimits_{k'\in\mathcal K}\operatorname{Tr}({\bf\Gamma}_k{\mathbf P}_{k'}{\bf\Gamma}_k^{\rm H}{\mathbf Q}^l) + \sigma^2)({\beta_k}-{\beta_k}^{[n_2-1]})).  
\end{aligned}    
\end{equation}
Hence, the solution of problem \eqref{SSE_9} can be iteratively optimized by  
solving the reconstructed problem:
\begin{subequations}\label{SSE_10}
\begin{align}
\max_{{\bf c, P},\beta_k}  &\quad t \tag{\ref{SSE_10}}\\
& t \le {\tilde f}\left(\tau_k^{[t]}, {\mathbf P}^{[t]}, {{\bf\Theta}^{[t]}}^l\right) - \lambda \phi({\bf P}), \label{SSE_10a} \\
& \tau_k = {\beta_k}^{\mu_k}, \label{SSE_10b} \\
\text { s.t. } &  \eqref{P_1_c5}, \eqref{P_2_in}-\eqref{P_2_e}, \eqref{com_strm_cons}.  \nonumber
\end{align}
\end{subequations} 
After obtaining $\{{\bf P}_m^{[n]},m\in\tilde{\mathcal K}\}$, the associated rank-one beamforming solution $\{{\bf w}_m^{*},m\in\tilde{\mathcal K}\}$ can be recovered via Gaussian randomization procedure \cite{Wang2014SDR}. Algorithm \ref{alg:proposed} summarizes the overall optimization procedure.
\begin{algorithm}[htbp]
\caption{Proposed Iterative Algorithm}
\label{alg:proposed}
\begin{algorithmic}[1]
\Require $\mathbf{H},\{\mathbf{h}_k\},\{\mathbf{g}_j\},P_{\max},r_0,r_{\mathrm{E}},\mu_k,\epsilon$
\Ensure $\{\mathbf{w}_m^*\},{\mathbf{\Theta}^*}^{t},{\mathbf{\Theta}^*}^{r}$, the optimal rate $R^*$
\State \textbf{Initialize:} $n\gets0$, $\{\mathbf{P}_m^{[0]}\}$, ${{\bf Q}^{[0]}}^l,{{\bf Q}^{[0]}}^r$, $\lambda^{[0]}$, $\tau_k^{[0]}$, $R^{[0]}$
\Repeat
\State $n\gets n+1$
\State \textbf{Phase shift optimization:}
\State $n_1\gets 0$, ${{\bf Q}^{[n_1]}}^l \leftarrow {{\bf Q}^{[n-1]}}^l$
\Repeat
\State $n_1 \gets n_1+1$
\State $\rho = 10^{(n_1-5)}$
\State Solve (12) with $\{\!\tau_k^{[n\!-\!1]}\!, \!\mathbf{P}_m^{[n\!-\!1]}\!, \!\lambda^{[n\!-\!1]}\}\!$ to get ${{\bf Q}^{[n_1\!]}}^l$
\Until{$\max_{l\in\{t,r\}}(\|{{\bf Q}^{[n_1]}}^l\|_*-\|{{\bf Q}^{[n_1-1]}}^l\|_2)\le\epsilon$}
\State ${{\bf Q}^{[n]}}^l\leftarrow{{\bf Q}^{[n_1]}}^l$, $l\in\{r,t\}$
\State \textbf{Beamforming and rate allocation:}
\State $n_2\gets0$, $\{{{\bf P}_m^{[n_2]}}\} \leftarrow \{{{\bf P}_m^{[n-1]}}\}$
\Repeat
\State $n_2\gets n_2+1$
\State Compute (14) and (16) around $\{\mathbf{P}_m^{[n_2-1]}\}$
\State Solve (17) to get $\{\mathbf{P}_m^{[n_2]},\tau_k^{[n_2]}\}$
\Until{$|t^{[n_2]}-t^{[n_2-1]}|\le\epsilon$}
\State $\{\mathbf{P}_m^{[n]}\}\leftarrow\{\mathbf{P}_m^{[n_2]}\}$, $\tau_k^{[n]}\leftarrow\tau_k^{[n_2]}$
\State $\lambda^{[n]}\leftarrow\min_{k\in\mathcal{K}}\frac{\tau_k^{[n]}\psi_k(\mathbf{P}^{[n]},{\mathbf{\Theta}^{[n]}}^l)}{\phi_k(\mathbf{P}^{[n]},{\mathbf{\Theta}^{[n]}}^l)}$
\Until{$|R^{[n]}-R^{[n-1]}|\le\epsilon$}
\State Recover $\{\mathbf{w}_m^*\}$ from $\{\mathbf{P}_m^{[n]}\}$ \State \Return $\{\mathbf{w}_m^*\}$, ${\mathbf{\Theta}^*}^{t},{\mathbf{\Theta}^*}^{r}$, $R^*=\min_k R_{k,\mathrm{tot}}^{[n]}$
\end{algorithmic}
\end{algorithm}

Finally, we analyze the convergence and computational complexity of the proposed algorithm. The objective function \eqref{P_2} is monotonically non-decreasing over iterations and is lower-bounded by the limited transmit power. As the penalty terms successively decrease during phase shift updates, the proposed algorithm is guaranteed to converge to a stationary point of the original problem when the penalty terms approach zero \cite{shi2020penalty}. 
The computational complexity per iteration is primarily determined by solving  \eqref{SSE_26} and \eqref{SSE_10} using interior-point method, whose respective time complexity are $\mathcal{O}(I_1 (2N_S)^{3.5})$ and $\mathcal{O}(I_2 (K N_T)^{3.5})$, where $I_1$ and $I_2$ denote iteration numbers of the respective subproblems.
Accordingly, the overall complexity is $\mathcal{O}\!\left(I\big(I_1 (2N_S)^{3.5}+I_2 (K N_T)^{3.5}\big)\log(\epsilon^{-1})\right)$, where $I$ denotes the number of outer iterations and $\epsilon$ is the convergence accuracy.
The space complexity is mainly determined by the storage of the optimization variables and channel matrices, which requires $\mathcal{O}((K+1)N_T^2+2N_S^2)$ and $\mathcal{O}(N_SN_T+KN_S+JN_S)$ memory. The overall space complexity is on the order of $\mathcal{O}(KN_T^2 + N_S^2)$.
Therefore, the time and space complexity grows polynomially with the number of STAR-RIS elements $N_S$, the number of transmit antennas $N_T$, and the number of users $K$. Although the computational cost increases for very large-scale systems, the proposed algorithm remains tractable for moderate network sizes typically considered in STAR-RIS-assisted networks.

\section{Numerical Results}
We validate the performance of the proposed RSMA-STAR scheme through simulations.
The transmitter has $N_T=8$ antennas and is located 50 meters from the STAR-RIS, equipped with $N_S=20$ elements, at an angle of $20^\circ$.
Channels follow Rician fading \cite{wang2022coupled}.
Unless otherwise specified, users and eavesdroppers are randomly located on a semicircle of radius $d = 80$m centered at the STAR-RIS.
The transmit SNR is ${\rm SNR} = \frac{P_{\rm max}}{\sigma^2} = 10$dB.
Set $r_0=r_{\rm E}$ so that all receivers have identical decoding capability.
Four benchmarks are considered: SDMA (No common stream), RSMA-RIS (STAR-RIS replaced by adjacent reflect-only  and  transmit-only RISs with $\frac{N_S}{2}$ elements), RSMA upperbound (security constraints ignored), and RSMA-Random (random  STAR-RIS phase shifts). 

\begin{figure}[!t]
\centering
\includegraphics[width=2.5in]{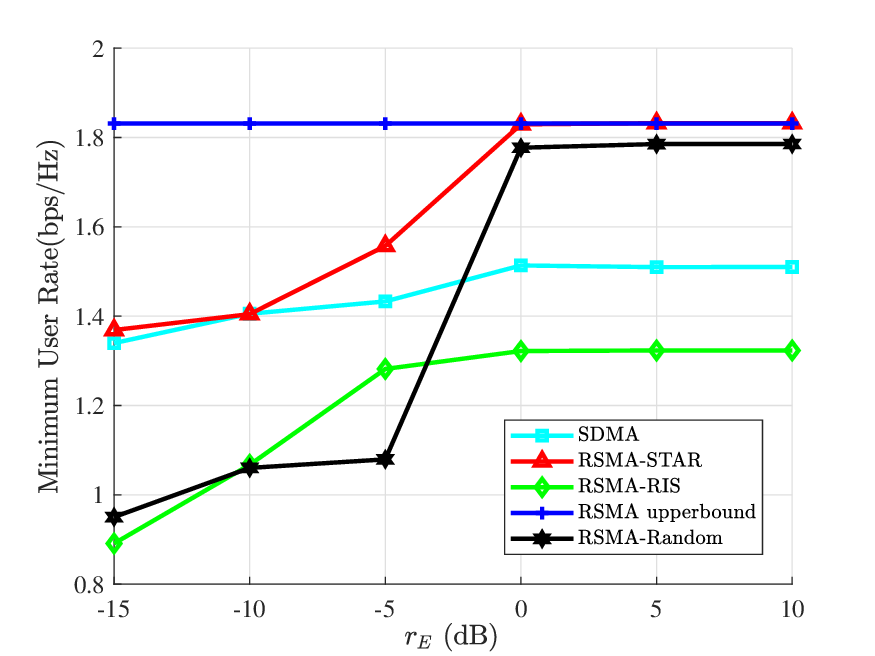}
\caption{Minimum user rate versus the secrecy threshold $r_{\rm E}$.}
\label{fig2}
\end{figure}
Fig. \ref{fig2} shows the minimum user rate versus the secrecy threshold $r_{\rm E}$.  
As secrecy requirement becomes tighter, all rates decrease due to stricter SINR constraints at eavesdroppers. 
The proposed RSMA-STAR scheme achieves the highest rate with a negligible gap to the upperbound when $r_{\rm E}\ge 0$dB, because the common stream can flexibly serve as artifical noise against external eavesdroppers while private streams are jointly optimized to avoid information leak to internal eavesdroppers, demonstrating an efficient balance between confidentiality and spectral efficiency.
The RSMA–RIS and RSMA–Random suffer noticeable degradation, highlighting the advantage of STAR-RIS and joint optimization.

\begin{figure}[!t]
\centering
\includegraphics[width=2.5in]{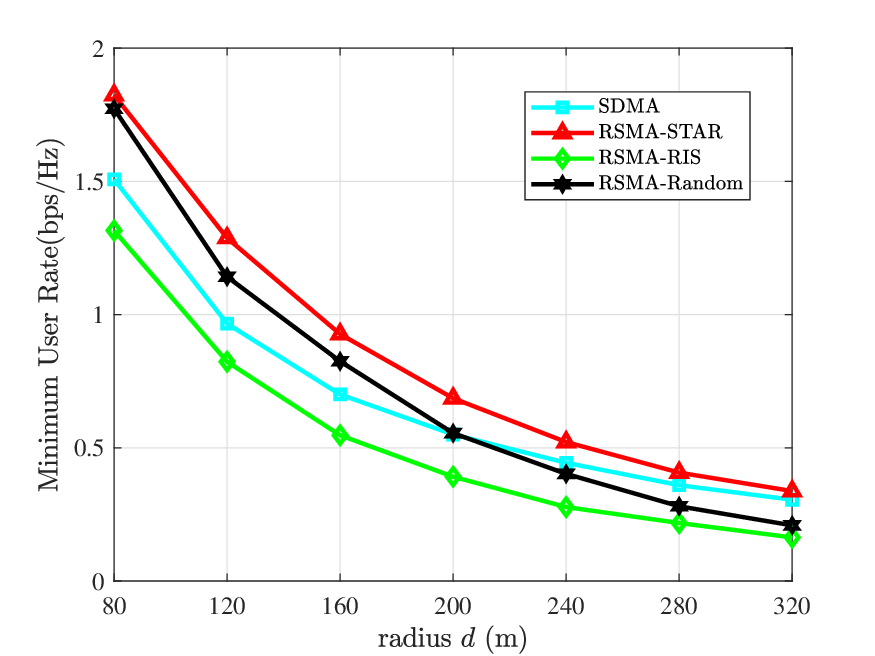}
\caption{Minimum user rate versus the distribution radius $d$,  $r_{\rm E}=0$dB.}
\label{fig5}
\end{figure} 
Fig. \ref{fig5} depicts the minimum user rate versus the distribution radius $d$. 
All rates decline with larger $d$ due to increased path loss. The RSMA–STAR scheme outperforms SDMA and RSMA–RIS by over 20\% when $d=80$m, owing to STAR-RIS’s full-space coverage and RSMA's efficient interference management. Moreover, the performance gap between RSMA-STAR and SDMA narrows with increasing $d$, while that over RSMA-Random widens, indicating that phase-shift optimization becomes more crucial compared to beamforming optimization at larger distances.

\begin{figure}[!t]
\centering
\includegraphics[width=2.5in]{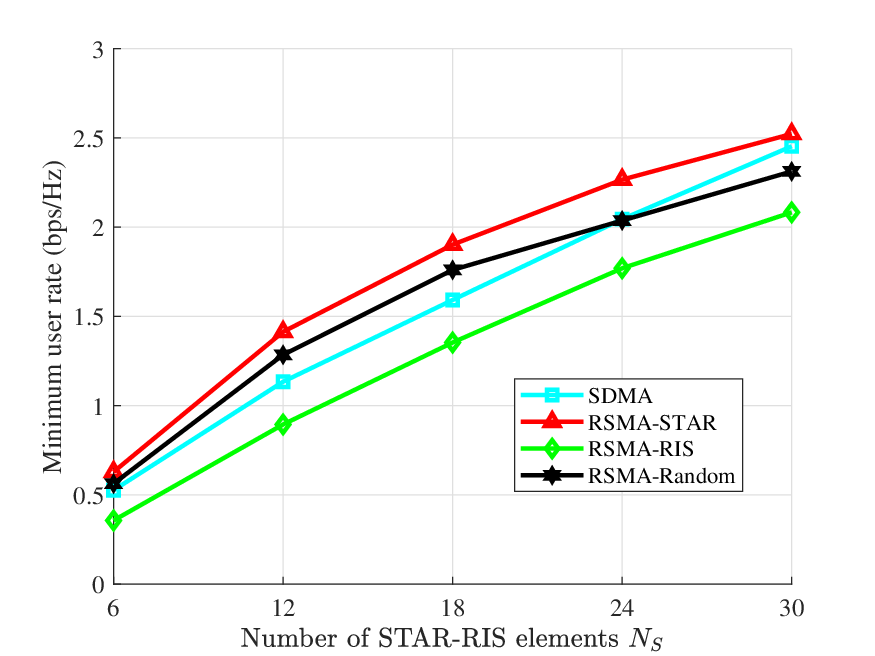}
\caption{Minimum user rate versus the number of STAR-RIS elements $N_{\rm S}$.}
\label{fig6}
\end{figure}
Fig. \ref{fig6} demonstrates the minimum user rate versus the number of STAR-RIS elements $N_{\rm S}$.
The rate increases steadily with $N_{\rm S}$ for all schemes. 
Throughout the range of $N_{\rm S}$ from 6 to 30, the proposed algorithm consistently outperforms the baselines, validating its robustness to varied STAR-RIS sizes.

\begin{figure}[!t]
\centering
\includegraphics[width=2.5in]{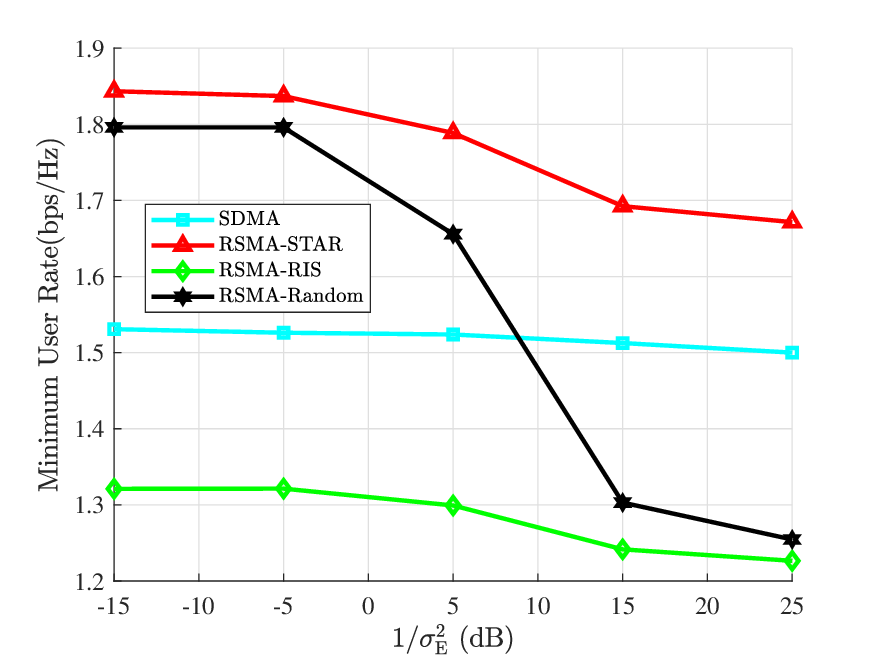}
\caption{Minimum user rate vs the noise variance ratio $1/\sigma_{\rm E}^2$.}
\label{fig4}
\end{figure}
Fig. \ref{fig4} shows the minimum user rate versus the noise variance ratio $1/\sigma_{\rm E}^2$. 
As $1/\sigma_{\rm E}^2$ increases, the noise variance at eavesdroppers decreases,  enhancing their wiretapping capabilities \cite{Liao2011QoS}.
The RSMA-STAR scheme exhibits strong robustness against noise uncertainty. 
As $1/\sigma_{\rm E}^2$ increases, the gap to SDMA narrows while that to RSMA-RIS remains constant, indicating reduced power allocated to the common stream, since the rate gain of RSMA-STAR over SDMA primarily stems from the common stream. This suggests that when the wiretap channel improves, the common stream power should be lowered to prevent decoding by external eavesdroppers. Moreover, the fixed equal common rate allocation limits adaptability under stronger eavesdropping conditions and a more flexible allocation is left for future work.

\section{Conclusion}
We studied secure STAR-RIS-assisted RSMA communications in the presence of both internal and external eavesdroppers. A max-min fairness problem was formulated and solved by an efficient iterative algorithm that jointly optimizes beamforming and phase shifts. 
Numerical results indicate that the proposed design achieves a favorable trade-off between spectral efficiency and secrecy.
Notably, the performance advantage remains stable with increasing STAR-RIS elements but decreases with better eavesdropper channel conditions.
Future work will consider more realistic extensions by accounting for imperfect CSI, direct-link modeling and imperfect SIC. 

\bibliographystyle{IEEEtran}
\bibliography{IEEEabrv,sRSMA_STAR-RIS}

@STRING{IEEE_J_VT         = "{IEEE} Trans. Veh. Technol."}

@STRING{IEEE_J_SP         = "{IEEE} Trans. Signal Process."}

@STRING{IEEE_J_COML       = "{IEEE} Commun. Lett."}

@STRING{IEEE_J_WCOM       = "{IEEE} Trans. Wireless Commun."}

@STRING{IEEE_J_WCOML      = "{IEEE} Wireless Commun. Lett."}

@STRING{IEEE_J_TGCN       = "{IEEE} Trans. Green Commun. Networking"}

@STRING{IEEE_J_IOT        = "{IEEE} Internet Things J."}

@STRING{IEEE_O_CSTO       = "{IEEE} Commun. Surveys Tuts."}

@STRING{ieee="IEEE Computer Society Press"}

@ARTICLE{Xu2021STAR-RIS,
  author={Xu, Jiaqi and Liu, Yuanwei and Mu, Xidong and Dobre, Octavia A.},
  journal=IEEE_J_COML, 
  title={{STAR-RISs}: Simultaneous Transmitting and Reflecting Reconfigurable Intelligent Surfaces}, 
  year={Sept. 2021},
  volume={25},
  number={9},
  pages={3134-3138},
  doi={10.1109/LCOMM.2021.3082214}}

@ARTICLE{Zhou2024STARjamming,
  author={Zhou, Tao and Xu, Kui and Hu, Guojie and Xia, Xiaochen and Xie, Wei and Li, Chunguo},
  journal=IEEE_J_TGCN, 
  title={Robust Beamforming Design for {STAR-RIS}-Assisted Anti-Jamming and Secure Transmission}, 
  year={Mar. 2024},
  volume={8},
  number={1},
  pages={345-361},
  doi={10.1109/TGCN.2023.3329127}}

@ARTICLE{Zargari2021MMF,
  author={Zargari, Shayan and Khalili, Ata and Wu, Qingqing and Robat Mili, Mohammad and Ng, Derrick Wing Kwan},
  journal=IEEE_J_VT, 
  title={Max-Min Fair Energy-Efficient Beamforming Design for Intelligent Reflecting Surface-Aided {SWIPT} Systems With Non-Linear Energy Harvesting Model}, 
  year={Jun. 2021},
  volume={70},
  number={6},
  pages={5848-5864},
  doi={10.1109/TVT.2021.3077477}}

@ARTICLE{Xia2024sRSMA,
  author={Xia, Huiyun and Mao, Yijie and Zhou, Xiaokang and Clerckx, Bruno and Han, Shuai and Li, Cheng},
  journal=IEEE_J_WCOM, 
  title={Weighted Sum-Rate Maximization of Rate-Splitting Multiple Access with Confidential Messages}, 
  year={Oct. 2024},
  volume={23},
  number={10},
  pages={13738-13751},
  doi={10.1109/TWC.2024.3404095}}

@ARTICLE{hao2020robustsecureRS,
  author={Fu, Hao and Feng, Suili and Tang, Weijun and Ng, Derrick Wing Kwan},
  journal=IEEE_J_WCOM, 
  title={Robust Secure Beamforming Design for Two-User Downlink {MISO} Rate-Splitting Systems}, 
  year={Dec. 2020},
  volume={19},
  number={12},
  pages={8351-8365},
  doi={10.1109/TWC.2020.3021725}}

@article{XIAO2023PLS,
title = {Physical layer security of {STAR-RIS-aided RSMA} systems},
journal = {Phys. Commun.},
volume = {61},
pages = {102192},
year = {Dec. 2023},
issn = {1874-4907},
author = {Fengcheng Xiao and Pengxu Chen and Shuobo Xu and Xiyu Pang and Hongwu Liu},
}

@ARTICLE{Hashempour2024secureSTAR,
  author={Hashempour, Hamid Reza and Bastami, Hamed and Moradikia, Majid and Zekavat, Seyed A. and Behroozi, Hamid and Berardinelli, Gilberto and Swindlehurst, A. Lee},
  journal=IEEE_J_VT, 
  title={Secure {SWIPT} in the Multiuser {STAR-RIS} Aided {MISO} Rate Splitting Downlink}, 
  year={Jun. 2024},
  volume={73},
  number={9},
  pages={1-16},
  doi={10.1109/TVT.2024.3398057}}

@ARTICLE{Zhang2024secureSTAR,
  author={Zhang, Yiran and Yang, Liang and Li, Xingwang and Guo, Kefeng and Liu, Hongwu},
  journal=IEEE_J_IOT, 
  title={Covert Communications for {STAR-RIS}-Assisted Industrial Networks With a Full Duplex Receiver and {RSMA}}, 
  year={Jun. 2024},
  volume={11},
  number={12},
  pages={22483-22493},
  doi={10.1109/JIOT.2024.3381166}}

@ARTICLE{Wang2014SDR,
  author={Wang, Kun-Yu and So, Anthony Man-Cho and Chang, Tsung-Hui and Ma, Wing-Kin and Chi, Chong-Yung},
  journal=IEEE_J_SP, 
  title={Outage Constrained Robust Transmit Optimization for Multiuser {MISO} Downlinks: Tractable Approximations by Conic Optimization}, 
  year={Nov. 2014},
  volume={62},
  number={21},
  pages={5690-5705},
  doi={10.1109/TSP.2014.2354312}}

@ARTICLE{Liao2011QoS,
  author={Liao, Wei-Cheng and Chang, Tsung-Hui and Ma, Wing-Kin and Chi, Chong-Yung},
  journal=IEEE_J_SP, 
  title={{QoS}-Based Transmit Beamforming in the Presence of Eavesdroppers: {An} Optimized Artificial-Noise-Aided Approach}, 
  year={Nov. 2011},
  volume={59},
  number={3},
  pages={1202-1216},
  doi={10.1109/TSP.2010.2094610}}

@ARTICLE{Xie2025STAR,
  author={Xie, Tingyu and Nan, Guoshun and Cui, Qimei and Deng, Gang and Du, Haitao and Tao, Xiaofeng},
  journal=IEEE_J_COML, 
  title={{Joint optimization of active and passive beamforming for RIS-assisted secure communication under multiple eavesdroppers}}, 
  year={Aug. 2025},
  volume={29},
  number={8},
  pages={1943-1947},
  doi={10.1109/LCOMM.2025.3581052}}

@article{wang2022coupled,
  title={Coupled Phase-Shift {STAR-RIS}s: A General Optimization Framework},
  author={Wang, Zhaolin and Mu, Xidong and Liu, Yuanwei and Schober, Robert},
  journal=IEEE_J_WCOML,
  volume={12},
  number={2},
  pages={207--211},
  year={Feb. 2023},
  publisher={IEEE}
}

@ARTICLE{shi2020penalty,
  author={Shi, Qingjiang and Hong, Mingyi},
  journal=IEEE_J_SP, 
  title={Penalty Dual Decomposition Method for Nonsmooth Nonconvex Optimization—{Part I}: Algorithms and Convergence Analysis}, 
  year={Jun. 2020},
  volume={68},
  number={},
  pages={4108-4122},
  doi={10.1109/TSP.2020.3001906}}

@ARTICLE{Mao2022survey,
  author={Mao, Yijie and Dizdar, Onur and Clerckx, Bruno and Schober, Robert and Popovski, Petar and Poor, H. Vincent},
  journal=IEEE_O_CSTO, 
  title={Rate-Splitting Multiple Access: {Fundamentals}, Survey, and Future Research Trends}, 
  year={4th Quart.,2022},
  volume={24},
  number={4},
  pages={2073-2126},
  doi={10.1109/COMST.2022.3191937}}

\end{document}